\documentclass[3p,times]{elsarticle}

\usepackage{ecrc}

\usepackage{amsmath}
\usepackage{amssymb}

\newcommand{\rhoB}{\rho_{\scriptscriptstyle B}}

\usepackage[normalem]{ulem}  
\usepackage[dvips]{color} 
\newcommand{\comment}[1]{}

\renewcommand\sout{\bgroup \color{red} \ULdepth=-.5ex \ULset}

\volume{00}

\firstpage{1}

\journalname{Nuclear Physics A}

\runauth{T. Tsubakihara, A. Ohnishi}

\jid{NUPHA}

\jnltitlelogo{Nuclear Physics A}

\CopyrightLine{2012}{Published by Elsevier Ltd.}

\usepackage{amssymb}

\usepackage[figuresright]{rotating}

\begin{document}

\begin{frontmatter}

\dochead{}



\title{Three-body couplings in RMF 
	\\ and its effects on hyperonic star equation of state\tnoteref{Report}}
\tnotetext[Report]{Report number: YITP-12-97}

%
%

\author[label1]{K.~Tsubakihara}
\author[label2]{A.~Ohnishi}

\address[label1]{Meme Media Laboratory, Hokkaido University.}
\address[label2]{Yukawa Institute for Theoretical Physics, Kyoto University.}

\begin{abstract}
We develop a relativistic mean field (RMF) model
with explicit three-body couplings
and apply it to hyperonic systems and neutron star matter.
%
%
Three-baryon repulsion is a promising ingredient to answer
the massive neutron star puzzle;
when strange hadrons such as hyperons are taken into account,
the equation of state (EOS) becomes too soft
to support the observed two-solar-mass neutron star.
We demonstrate that it is possible to consistently explain
the massive neutron star and hypernuclear data
when we include three-body couplings and modify the hyperon-vector meson
couplings from the flavor SU(3) value.
\end{abstract}


\end{frontmatter}


\section{Introduction}
\label{Sec:Intro}
In constructing dense matter equations of state (EOSs),
it is strongly desired to respect hypernuclear data;
hyperons are expected to emerge in the neutron star core,
and they drastically soften the dense matter EOS.
While nucleonic EOSs without hyperons predict
the maximum mass of neutron stars as $(1.5-2.7) M_\odot$,
hyperons are expected to appear and reduce the maximum mass
to $(1.3-1.6) M_\odot$~\cite{Lattimer}.
Contrary to these understandings,
a two-solar-mass neutron star ($M = 1.97 \pm 0.04 M_\odot$)
is discovered recently by using the Shapiro delay,
and it is claimed that most of the EOSs with the appearance of strange hadrons
are ruled out~\cite{Demorest}.
%
%
It is also questionable even for surviving nucleonic EOSs 
to support the massive neutron star, when hyperons are introduced.
Thus we need to find
either the reason why hyperons do not appear in dense neutron star matter
or the mechanism how EOSs can be stiff enough even with hyperons.

One of the possible mechanisms to make the EOS stiffer
is the three-baryon repulsion.
In microscopic G-matrix calculations,
the three-baryon repulsion is found to be necessary
to support the 1.44~$M_\odot$ neutron star
when hyperons are included~\cite{NTY}.
In relativistic mean field (RMF) models,
attractive contribution from the scalar potential
grows more slowly than the baryon density.
In terms of non-relativistic languages,
this behavior can be interpreted as the {\it implicit}
three-body repulsion
caused by relativistic effects.
This three-body repulsion had been considered
to be enough to support neutron stars 
even if hyperons are taken into account,
until the two-solar-mass neutron star was discovered.
When hyperon-meson couplings are chosen away from the SU(6) values
and the $\omega$ meson self-energy is ignored,
the calculated neutron star maximum mass can be compatible
with the observed massive neutron star~\cite{Weissenborn}.
However, these hyperon-meson couplings have not been seriously verified
in finite hypernuclear systems,
and the density dependence of the vector potential in these treatments
would not be compatible
with the relativistic Br{\"u}ckner-Hartree-Fock (RBHF) calculation~\cite{RBHF},
whose results ensure that RMF models are reasonable.
Thus the most natural way to make the EOS with hyperons stiff enough 
would be to introduce {\em explicit} three-body repulsion.

In this work, we examine how 
three-body couplings affect the neutron star matter EOS
in the framework of RMF.
We include the interaction terms
of baryon-meson-meson ($BMM$) and three-meson ($MMM$) couplings.
Two mesons in the $BMM$ term couple with two other baryons,
and three mesons in the $MMM$ term couple with three baryons.
Thus $BMM$ and $MMM$ couplings correspond to the explicit three-body forces.
Each term in the RMF Lagrangian can be characterized
by the number $n=B/2+M+D$
in the Furnstahl-Serot-Tang (FST) expansion~\cite{FST},
where $B$ is the number of baryon fields,
$M$ is the number of non-Nambu-Goldstone boson fields,
and $D$ is the number of derivatives.
The baryon-meson coupling terms $\bar{B}MB$ belong to $n=2$,
and the present three-body coupling terms correspond to $n=3$
---
the next-to-leading order interactions in the FST expansion.
While some of the $n=3$ terms are considered to be absorbed
by field redefinitions~\cite{FST},
we need to modify higher order ($n\geq4$) terms
to compensate the redefinitions.
We demonstrate the importance of $n=3$ terms,
especially the $BMM$ terms, on the dense matter EOS.

\section{Model description}

We adopt here an RMF Lagrangian,
$\mathcal{L}_{\mbox{\scriptsize RMF}}
 = \mathcal{L}_{n=2} + V_{\mbox{\scriptsize eff}} + \mathcal{L}_{n=3}$,
where $\mathcal{L}_{n=2}$ corresponds to
the ordinary RMF Lagrangian including $n=2$ two-body couplings,
and $V_\mathrm{eff}$ represents the meson self-energies,
which include an $\omega^4$ potential and 
a logarithmic potential of scalar-isoscalar mesons~\cite{Tsubaki2010}.
Here we ignore the nucleon coupling with
scalar and vector $\bar{s}s$ mesons ($\zeta$ and $\phi$),
as usually assumed.
For the three-body coupling terms, $\mathcal{L}_{n=3}$,
we first consider symmetric nuclear matter case,
where $\mathcal{L}_{n=3}$ is taken to be
%
\begin{align}
\mathcal{L}_{n=3}^{\sigma\omega}
&=- \frac{1}{f_\pi}\sum_B \bar{\psi}_B \left[
 g_{\sigma\sigma{B}} \sigma^2
+g_{\omega\omega{B}} \omega_\mu\omega^\mu
-g_{\sigma\omega{B}} \sigma\omega_\mu\gamma^\mu
\right] \psi_B
-c_{\sigma\omega\omega} f_\pi\sigma\omega_\mu\omega^\mu
\ .\label{Eq:n=3}
\end{align}
The first and second terms modify the effective mass of nucleons,
$M_N^* = M_N - (g_{\sigma N}\sigma
-g_{\sigma\sigma N}\sigma^2/f_\pi-g_{\omega\omega N}\omega^2/f_\pi)$,
where $\omega$ represents the temporal component.
These $n=3$ terms modify the effective mass
from the $n=2$ coupling with $\sigma$.
The third term modifies the vector potential of nucleons at high density,
$U_v=(g_{\omega N}-g_{\sigma\omega N}\sigma/f_\pi)\omega$,
and the fourth term represents the $\omega$ meson mass shift at finite density.
In RBHF calculations, the vector potential at low densities
is almost proportional to the baryon density,
but the vector potential to baryon density
ratio $U_v/\rhoB$ is suppressed
around $\rho_0$ or at higher densities.
This suppression in RBHF is sometimes simulated
by the $\omega^4$ term~\cite{TM1}
or by the density dependent coupling~\cite{DDRMF}.
When we simulate the suppression only with the $\omega^4$ term,
the ratio $U_v/\rhoB$ is monotonically decreasing with increasing $\rhoB$.
With a large coefficient of $\omega^4$,
EOS at high density is thus too softened
to support the massive neutron star~\cite{Tsubaki2010}.
The density dependent vector coupling is usually chosen
to decrease and be saturated with increasing $\rhoB$.
With the present $n=3$ coupling terms in Eq.~\eqref{Eq:n=3}
in addition to the $\omega^4$ term,
we try to simulate the decreasing and saturating
behavior of the $U_v/\rhoB$ ratio found in the RBHF results.

For the massive neutron star puzzle,
hyperon-meson coupling is another key.
In many of the RMF parameter sets,
the hyperon- and nucleon-vector meson coupling ratio
is chosen to be 
$R \equiv g_{\omega Y}/g_{\omega N} \simeq 2/3 (Y=\Lambda, \Sigma)$
based on the spin-flavor SU(6) symmetry or the quark counting arguments.
This choice is the main reason why we cannot support the massive neutron 
star with hyperons.
Mesons in RMF models describe scalar and vector potentials
from various origins;
two pion exchange, correlation from two-baryon short range repulsion,
meson pair exchanges, and so on,
in addition to the meson fields consisting of $\bar{q}q$.
Thus it is not mandatory to impose the spin-flavor SU(6)
or flavor SU(3) relations among the coupling constants.
In our previous work~\cite{Tsubaki2010},
we have adopted a more phenomenological prescription;
while the hyperon-isoscalar vector meson couplings
($g_{\omega Y}, g_{\phi Y}$)
have been chosen to be the flavor SU(3) values,
other couplings in the $S=-1$ hyperon sector
($g_{\sigma Y}, g_{\zeta Y}, g_{\rho \Sigma}$)
have been fitted to the hypernuclear data,
including $\Lambda$ separation energies ($S_\Lambda$), 
the bond energy of the double $\Lambda$ hypernucleus
($\Delta B_{\Lambda\Lambda}$), and the $\Sigma^-$ atomic shift data.
It should be noted that
we need to adopt the $\Sigma$-$\rho$ coupling
much smaller than the flavor SU(3) value
in order to fit the $\Sigma^-$ atomic shift data~\cite{Tsubaki2010,Mares}.
We here explore the results using the hyperon-$\omega$ coupling value 
other than the SU(3) value.
Specifically, we examine the results with $R=0.8$ in the later discussion.

\section{Results}

\begin{figure}
	\centering
	\includegraphics[width=0.35\textwidth]{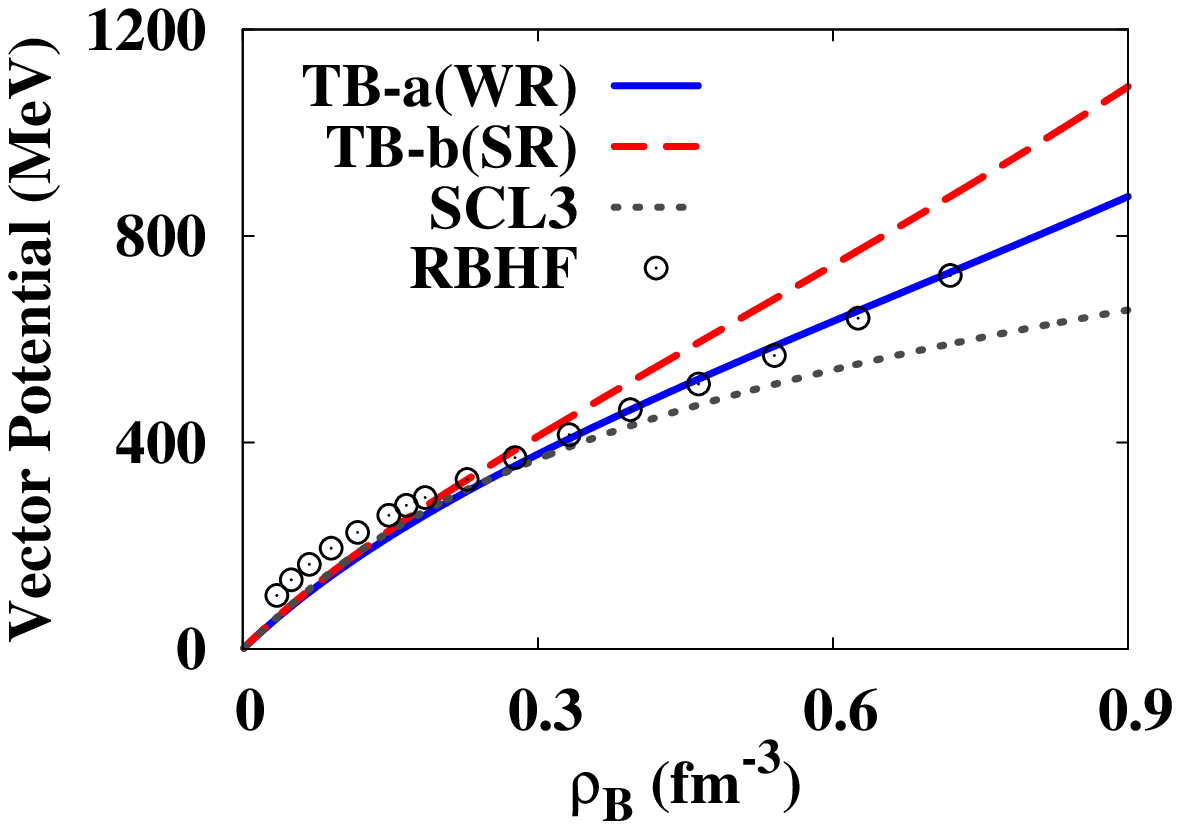}
	\hspace{20pt}
	\includegraphics[width=0.35\textwidth]{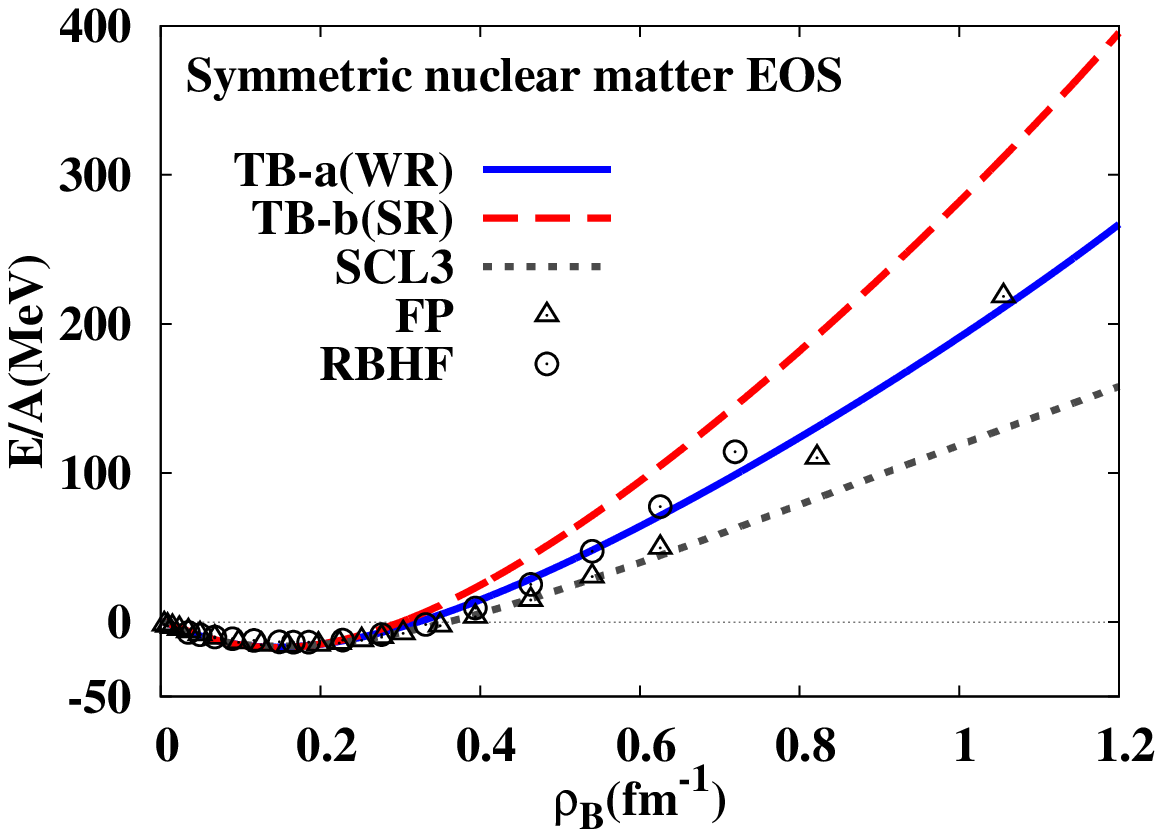}
	\caption{
		Calculated vector potential and EOS in symmetric nuclear matter
		with TB-a and TB-b parameter set
		in comparison with SCL3 results.
		}
	\label{Fig:SymEOS}
\end{figure}

We prepare two three-body coupling parameter sets, TB-a and TB-b.
The scalar meson part of $V_\mathrm{eff}$
is taken from our previous work
(SCL3 RMF model)~\cite{Tsubaki2010},
which describes known properties of finite and infinite nuclear systems
such as binding energies per nucleon ($B/A$).
Three-body couplings in TB-a are determined so as to reproduce
the density dependence of the vector potential in RBHF at high densities.
In TB-b, the repulsion from three-body couplings on the vector potential
is chosen to be about twice of that in TB-a
as shown in the left panel of Fig.~\ref{Fig:SymEOS}.
$n=2$ couplings are modified 
to reproduce the saturation point.
%
We adopt the parameters
$g_{\sigma N}=m_N/f_\pi$,
$g_{\omega N}=12.116(12.284)$,
$g_{\sigma\sigma N}=1.50(0.70)$,
$g_{\omega\omega N}=-1.50(-0.15)$,
$g_{\sigma\omega N}=3.05(3.45)$,
$c_{\sigma\omega\omega}=-3.0(-3.0)$,
and
$c_{\omega}=4.1015(-3.9186)$
for TB-a (TB-b), where $c_\omega$ is the coefficient
of $(\omega_\mu\omega^\mu)^2/4$.
%
Compared to symmetric nuclear matter EOS in SCL3,
EOSs in TB-a and TB-b are obviously stiffened
especially at high $\rho_B$ as shown in the right panel of
Fig.~\ref{Fig:SymEOS}.	
TB-a EOS shows good agreement with RBHF EOS and 
stiffer than Friedman-Pandharipande (FP) EOS~\cite{FP}.

\begin{figure}
	\centering
	\includegraphics[width=0.35\textwidth]{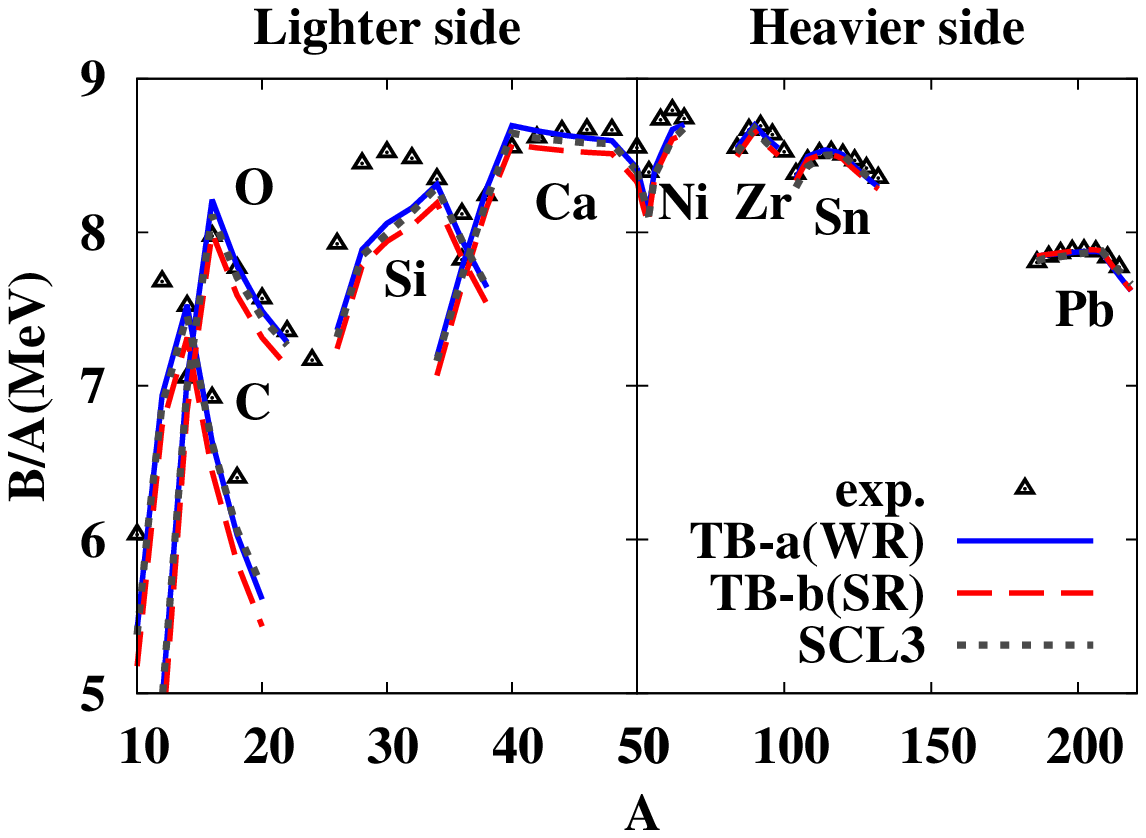}
	\hspace{20pt}
	\includegraphics[width=0.35\textwidth]{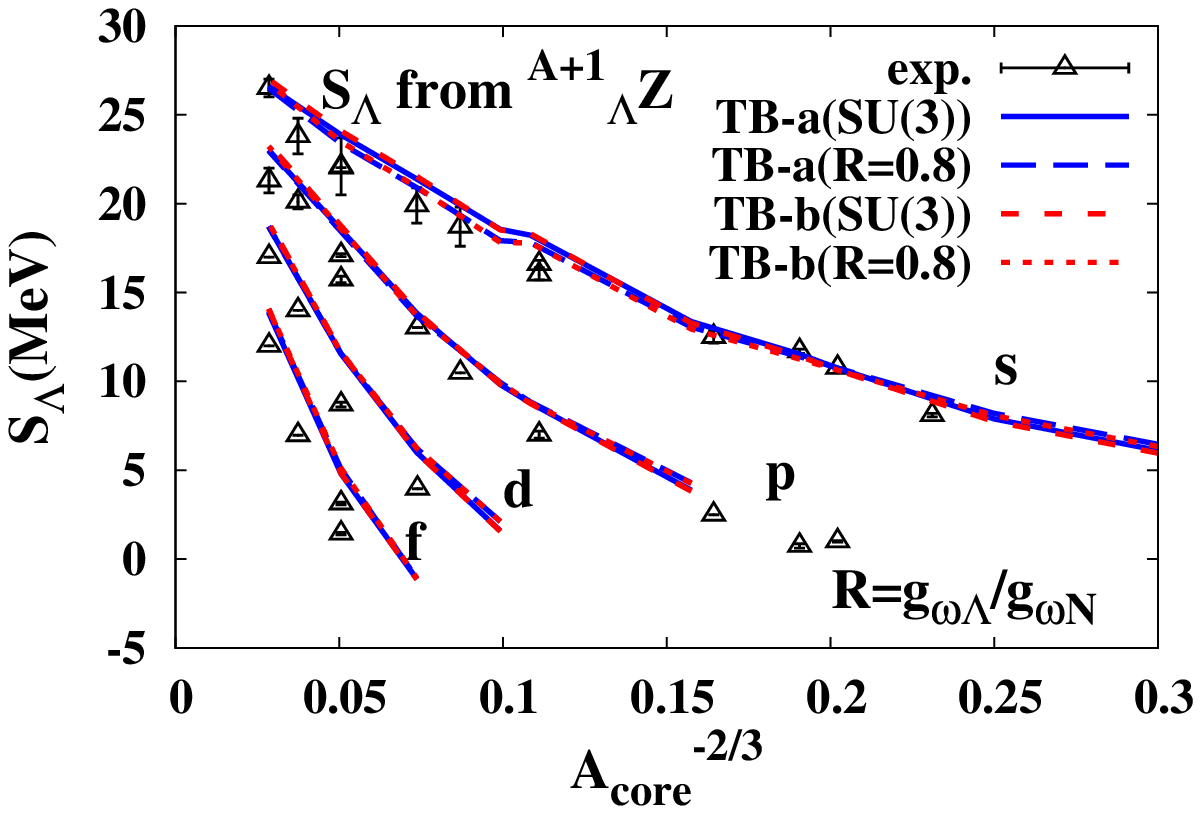}
	\caption{
		Calculated binding energies and $\Lambda$ separation energies
		($S_\Lambda$) based on TB-a and TB-b parameter sets.
		}
	\label{Fig:Finite}
\end{figure}

Isovector $n=2$ coupling and $\Lambda$-meson couplings are
obtained by fitting $B/A$, $S_\Lambda$ and $\Delta B_{\Lambda\Lambda}$.
With the SU(3) $\Lambda\omega$ coupling,
we choose 
$g_{\rho N}=4.86(4.87)$,
$g_{\sigma\Lambda}=3.46(3.45)$,
$g_{\zeta\Lambda}=4.73(4.71)$.
$g_{\sigma\sigma\Lambda}=0.50(0.23)$,
$g_{\omega\omega\Lambda}=-0.50(-0.05)$,
and
$g_{\sigma\omega\Lambda}=1.02(1.15)$
for TB-a (TB-b),
where $\Lambda MM$ couplings are weaker than nucleon cases.
In the case of
$R=0.8$,
$\Lambda$-scalar meson couplings are modified as
$g_{\sigma\Lambda}=4.71(4.73)$ and $g_{\zeta\Lambda}=5.16(5.19)$.
As shown in Fig.~\ref{Fig:Finite},
we can describe $B/A$ and $S_\Lambda$ reasonably well,
except for $B/A$ of light $jj$ closed nuclei.
We may introduce the tensor coupling of vector mesons
to obtain a larger $ls$ potential,
but the tensor coupling does not modify the uniform matter
EOS in the mean field treatment.
In the $R=0.8$ case, both scalar and vector potentials become stronger,
and we find slightly different trends in $S_\Lambda$
as shown in Fig.~\ref{Fig:Finite}.
%
$\Lambda$ single particle excitation
energies are calculated to be smaller than those without three-body couplings,
since the three-body couplings suppress the scalar potential and
$\Lambda$ effective mass is larger in the present parametrization.
It is necessary to tune hyperon-meson(-meson) couplings more carefully
in order to reproduce $\Lambda$ single particle energies,
especially of those in $d$- and $f$-shells in heavy hypernuclei.

\begin{figure}
	\centering
	\includegraphics[width=0.35\textwidth]{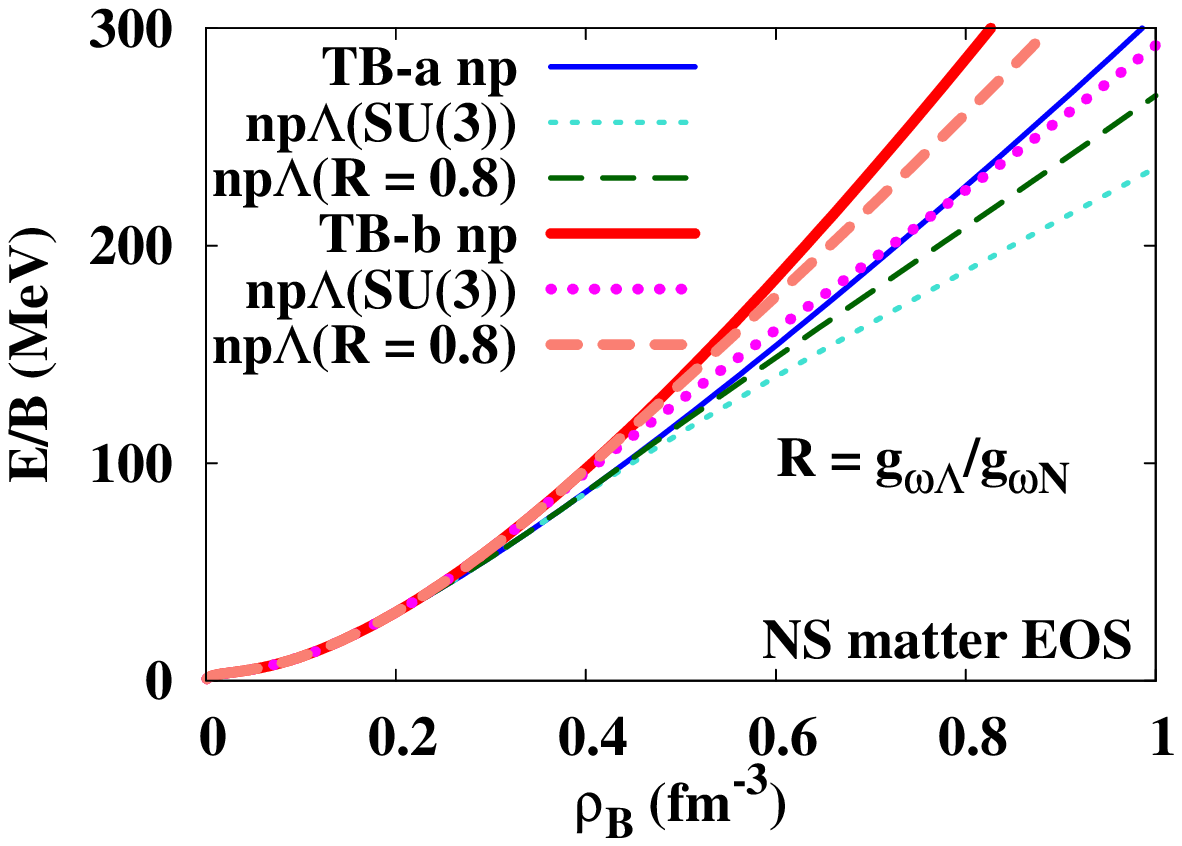}
	\hspace{20pt}
	\includegraphics[width=0.35\textwidth]{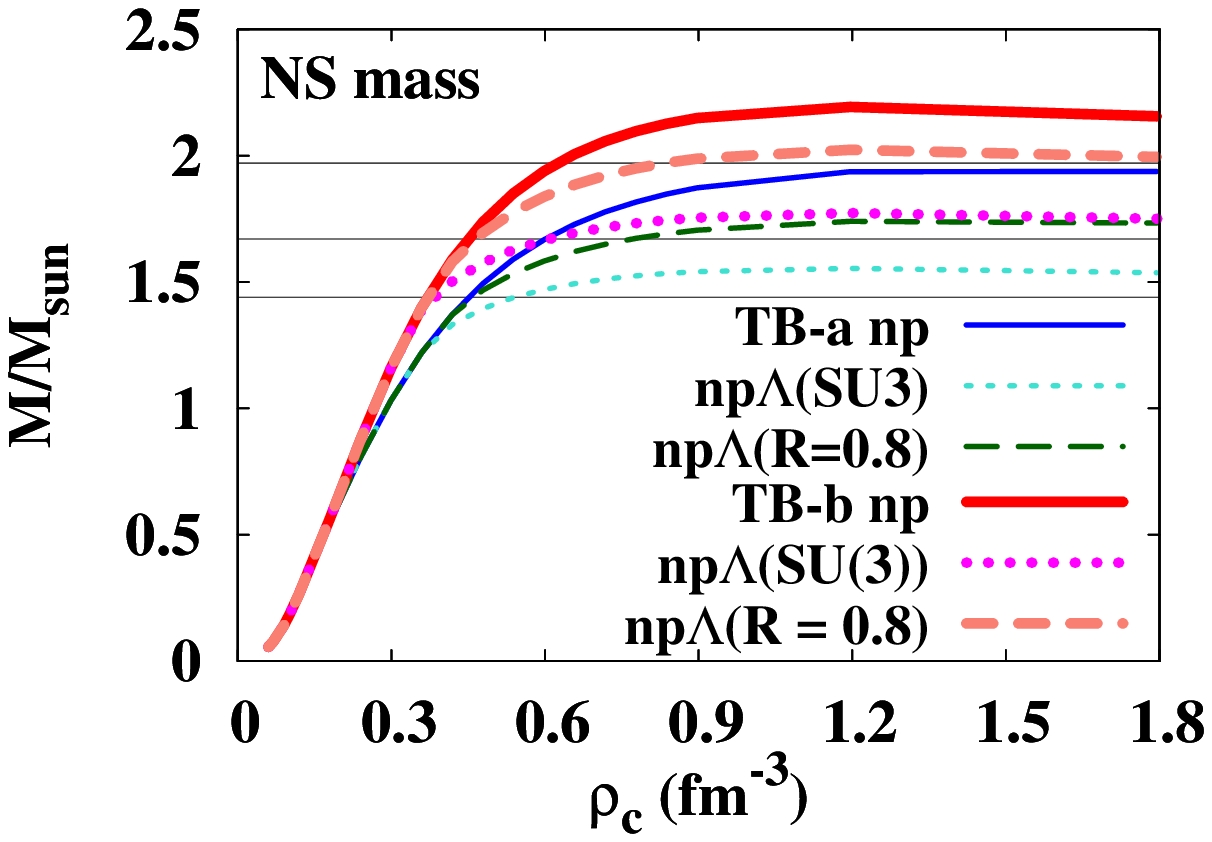}
	\caption{
		Neutron star matter EOSs as functions of $\rhoB$
		and neutron star maximum masses as functions of 
		the central density.
		}
	\label{Fig:NS}
\end{figure}

In Fig.~\ref{Fig:NS}, we show the neutron star matter EOS (NS EOS) obtained
with TB-a and TB-b.
By including three-body couplings, we can obtain stiffer NS EOSs
without spoiling the nuclear matter saturation and finite nuclear properties.
We also plot the mass-central density curve in the right panel
of Fig.~\ref{Fig:NS}.
We find that the calculated maximum mass exceeds $1.97 M_\odot$
by adopting repulsive three-body and hyperon-vector couplings
(TB-b and $R=0.8$),
while it seems to be difficult for TB-a to support the massive neutron star.

The above conclusion may not be in agreement with recent studies
using different type of higher-order terms~\cite{Bednarek:2011gd},
where the $2 M_\odot$ mass neutron star is found to be supported
in RMF with the SU(6) $R$ value.
%
One of the differences is the existence of the $\omega\rho$ coupling term.
The $\omega\rho$ coupling term
$\propto(\omega_\mu\omega^\mu)(\rho_\mu^a\rho^{a\mu})$
is related to the symmetry energy at high densities.
We have not included this coupling,
since it belongs to $n=4$ in the FST expansion.
We may need to include $n=4$ terms in addition to other $n=3$ terms
in order for a systematic study including the density dependence
of the symmetry energy.

\section{Summary and conclusion}

We have examined three-body couplings ($n=3$)
in the framework of the relativistic mean field (RMF) model.
We can obtain stiffer EOSs at high densities
while keeping the nuclear matter properties around $\rho_0$
and finite nuclei.
%
We have also demonstrated that
finite hypernuclear properties are reasonably well described
with a modified $g_{\omega \Lambda}$ from the flavor SU(3) value.
By these two modifications in the RMF Lagrangian,
we can obtain the neutron star matter EOSs,
which support the recently observed two-solar-mass neutron star.
These findings indicate that it would be possible to answer
the massive neutron star puzzle;
we can explain recently observed massive neutron star
even if we respect hypernuclear data.
The two ingredients discussed here
may afford a key to understand the maximum mass of neutron stars.
More careful tuning of parameters and introduction of other interaction terms
may be necessary for a more satisfactory description of finite 
normal nuclei and hypernuclei.
Isovector part of $n=3$ couplings should be also investigated,
and will be reported elsewhere.

\section*{Acknowledgments}
This work is supported in part by the Grants-in-Aid for Scientific Research
from JSPS
(Nos.
          (B) 23340067, 
          (B) 24340054, 
          (C) 24540271  
),
by the Grants-in-Aid for Scientific Research on Innovative Areas from MEXT
(No. 2404: 24105001, 24105008), 
by the Yukawa International Program for Quark-hadron Sciences,
and by the Grant-in-Aid for the global COE program ``The Next Generation
of Physics, Spun from Universality and Emergence" from MEXT.








\end{document}